\documentstyle[11pt] {article}

\title{A "Square-root" Method for the Density Matrix and its Applications to Lindblad Operators}
\author{A. Yahalom$^{a,b}$ and R. Englman$^{b,c}$\\
$^a$ Institute of Astronomy, University of Cambridge\\
Madingley Road, Cambridge CB3 0HA, United Kingdom\\
$^b$ College of Judea and Samaria, Ariel 44284, Israel\\
$^c$ Department of Physics and Applied Mathematics,\\
 Soreq NRC,Yavne 81800,Israel\\ \\
e-mail:  asya@yosh.ac.il; englman@vms.huji.ac.il;}

\begin{document}
\maketitle

\newcommand{\beq} {\begin{equation}}
\newcommand{\enq} {\end{equation}}
\newcommand{\ber} {\begin {eqnarray}}
\newcommand{\enr} {\end {eqnarray}}
\newcommand{\eq} {equation}
\newcommand{\eqs} {equations }
\newcommand{\mn}  {{\mu \nu}}
\newcommand{\sn}  {{\sigma \nu}}
\newcommand{\rhm}  {{\rho \mu}}
\newcommand {\SE} {Schr\"{o}dinger equation}
\newcommand{\sr}  {{\sigma \rho}}
\newcommand{\bh}  {{\bar h}}
\newcommand {\er}[1] {equation (\ref{#1}) }
\newcommand {\ern}[1] {equation (\ref{#1})}
\newcommand {\Er}[1] {Equation (\ref{#1}) }
\newcommand {\erb}[1] {equation (\ref{#1})}
\newcommand{\gb} {{\bf \gamma}}
\newcommand{\gcrb}  {{\bf \gamma^+}}
\newcommand{\gd} {{\dot \gamma}}
\newcommand{\gcr} {{\gamma^+}}
\newcommand{\gcrd} {{ \dot \gamma^+}}
\newcommand{\ro} {{ \gamma  \gamma^+}}
\newcommand{\gf} {{\gamma_1}}
\newcommand{\gs} {{\gamma_2}}

\begin {abstract}

The evolution of open systems, subject to both Hamiltonian and
dissipative forces, is studied by writing the $nm$ element of the
time ($t$) dependent density matrix in the form
\ber
\rho_{nm}(t)&=& \frac {1}{A} \sum
   _{\alpha=1}^A \gamma ^{\alpha}_n (t)\gamma^{\alpha *}_m (t) \nonumber
\enr
The so called "square root factors", the $\gamma(t)$'s,
are non-square matrices and are averaged
 over $A$ systems ($\alpha$) of the ensemble. This
square-root description is exact.  Evolution equations are then
postulated for the $\gamma(t)$ factors, such as to reduce to the
Lindblad-type evolution equations for the diagonal terms in the
density matrix. For the off-diagonal terms they differ from the
Lindblad-equations. The "square root factors" $\gamma(t)$ are not unique and the
equations for the $\gamma(t)$'s depend on the specific representation
chosen. Two criteria can be suggested for fixing the choice of $\gamma(t)$'s one
is simplicity of the resulting equations and the other has to do with the reduction
of the difference between the $\gamma(t)$ formalism and the Lindblad-equations.

  When the method is tested on cases which have been
 previously treated by other methods, our results agree with them.
 Examples chosen are ($i$) molecular systems, such that are either periodically driven
  near level degeneracies, for which we calculate the decoherence
occurring  in multiple Landau-Zener transition,  or else when undergoing
descent  around conical intersections in the potential surfaces, ($ii$) formal
dissipative systems with Lindblad-type operators representing
either a non-Markovian process or a two-state system coupled
to bosons.

Attractive features of the present factorization method are
complete positivity, the no higher than linear increase of
 the implementation effort with the number of states involved and
  the introduction of randomness only at the start of the process.

\end {abstract}
\\
\\
PACS number(s):~03.65.Bz

\section {Introduction}

The present authors have proposed a variational formulation to
study the time evolution  of the density matrix for situations
including both Hamiltonian and dissipative processes \cite
{EnglmanY2004}. This was based on an expression for the time
($t$)-dependent density matrix, originally devised in {\cite
{Reznik} and developed by \cite{SGS}, which was written formally as
\beq
\rho(t) = \gb(t)\cdot\gcrb(t)
\label{rho1}
\enq
 In this work we apply this "square root" or
"factorization" method to the investigation of quantum
trajectories in a decoherent environment.

In the above, the "square-root factors"  are the non-square matrix
 $\gamma(t) $ whose components are written as \beq
\gamma^{\alpha}_n(t)\label{gam1}\enq and its hermitean conjugate.
 The upper index $\alpha$
designates the system in the ensemble and  the
lower index $n$ the state of the system.  In principle, both labels
run over infinite values, but for bookkeeping purposes we let
$\alpha$ take $A$ values and $n$ take $N$ values. (Also, in the examples worked
out in this paper we have taken for $A$ and $N$ finite and small integers.) Thus, $\gamma$ the
primary quantity in the formalism is a rectangular $N$ x $A$
matrix. As derived in \cite {Band} and in other texts following von
Neumann's method \cite {Neumann}, the density matrix is obtained
  as an ensemble average over all systems. A general  $nm$ component is
   \beq
   \rho_{nm} = \frac {1}{A} \sum
   _{\alpha=1}^A \gamma^{\alpha}_n\gamma^{\alpha *}_m
   \label {rhomatrix1}
   \enq
[\Er{rhomatrix1} is quoted by numerous quantum mechanics and
statistical physics textbooks. The relationship of the $\gamma$'s
 to the wave-functions in an ensemble and other properties of the
 density matrix are given in the appendix. In \er{rho1} the
dot -symbol is a shorthand for the averaging
(a normalized inner product) over $\alpha$.]

    A formally identical form is obtained for $\rho$ from an alternative
    definition of the density matrix of an open system, namely by starting
    with  the total density matrix of the system ($s$) $+$  its interactive
   environment ($e$) and then taking the trace over the degrees of freedom
   of the environment \cite {Blum}. Thus the "factorization" of the density matrix involves
   no approximation.

   To show this, we write a state vector of the combined system as
   \beq
   |\Psi>_{s+e}=\sum_{n\alpha}g^{\alpha}_n|n>_{s} |\alpha >_{e}\label{Psi1}
   \enq
      in terms of complete sets of the system and of the environment. From this we form the
   density matrix operator
   \beq
   \hat{\rho}_{s+e}= |\Psi>_{s+e} <\Psi|_{s+e} =
   \sum_{n \alpha m \beta} g^{ \alpha}_n g^{\beta *}_m
   |\alpha>_e|n>_{s}<m|_{s}<\beta|_{e}
   \label{rhotot}
   \enq
   Tracing over the environment states
   gives the reduced density matrix for the system alone
   \beq
   \hat{\rho}_s={\bf Tr}_e \hat{\rho}_{s+e} =
   \sum_{\lambda} <\lambda|\hat{\rho}_{s+e}|\lambda> =
   \sum_{nm\alpha}g^{\alpha}_n  g^{\alpha *}_m |n>_{s}<m|_{s}
   \label {rhos}
   \enq
The $nm$ matrix element of the operator $\hat{\rho}_s$ is: \beq
\rho_{nm}=<n|\hat{\rho}_s|m> =\sum_{\alpha}g^{\alpha}_n g^{\alpha
*}_m \enq which is of the square root form in \er{rhomatrix1} with
$\gamma^{\alpha}_n=\sqrt{A}g^{\alpha}_n$.

\subsection{The non uniqueness of square-root factors }
\label{unique}

The square-root factors are not unique. This can easily be seen from \er{rho1}. Suppose I
know a matrix $\gb(t)$ and use it to calculate $\rho(t)$:
\beq
\rho(t) = \gb(t)\cdot\gcrb(t)
\label{rho1z}
\enq
Obviously I can take any unitary matrix $U$ in the ensemble space and use it to calculate another
matrix $\gb'(t)$ such that:
\beq
\gb'(t) = \gb(t) U
\label{gbp}
\enq
Or in index notation:
\beq
\gb'^{\alpha}_n (t) = \gb^{\beta}_n (t) U^{\alpha}_{\beta}
\label{gbpindex}
\enq
Now if $\gb'(t)$ is used to calculate another density matrix $\rho'(t)$ using \er{rho1} we obtain:
\beq
\rho'(t) = \gb'(t)\cdot\gcrb'(t) = \gb(t)U \cdot U^\dag \gcrb(t) =\gb(t)U \cdot U^\dag \gcrb(t)= \rho (t)
\label{rho1p}
\enq
Hence $\gb'(t) = \gb(t) U$ is just another representation of the density matrix $\rho (t)$, the
use of either $\gb'(t)$ or $\gb(t)$ can not be distinguished experimentally and has no physical significance.
Furthermore we can choose a $U$ matrix evolving in time such that $U = U(t)$.
The considerations that dictate which is the best representation to use are discussed
in the following sections.

\section{Evolution in a dissipative environment}

\subsection{A density matrix formulation}
At the present the Liouville-von Neumann-Lindblad (LvNL)
equations, that are linear in the density
 matrix and ensure its complete positivity, are frequently employed for the
 evolution of the density matrix in the presence of dissipative processes. These are written
 (with the over dot representing time derivation and the time dependence temporarily
 suppressed) as:
 \beq
 \hbar \dot{\rho}=-i[H,\rho] +(2L\rho L^+ - L^+ L\rho-\rho
L^+L) \label{Lindblad} \enq showing the Hamiltonian ($H$) and
dissipative ($L$) processes \cite {Lindblad}. (There may be
several of the latter, in which case each process is labelled by
an index and the rate equation contains a sum over the processes.
For notational simplicity we consider a single process and do
without an indexed $L$. A theoretical development leading to the
above equations is found in \cite {Louisell}.) The relation of the
Lindblad equation to stochastic (Ito or Stratonovich) formulation
of dissipative processes in quantum mechanics was developed in
\cite {GisinP}, further elucidated in \cite{WisemannM} and
comparisons between various rate formalism were made in \cite
{KohenMT}. Several extensions of the Monte-Carlo (MC) or
unravelling formalism were made to non-Markovian and other
processes (e.g., \cite {StrunzDG}-\cite{Budini}). For opposing
view points we refer to \cite {Weiss}-\cite{RomeroTH}.

\subsection {Time development of the $\gamma$'s}

 We now postulate time-evolution equations  in the "square root" method. The form
 of the equations is rationalized by the facts that
  they will correctly give the Liouville-von Neumann equations for Hamiltonian
  processes and that they have the form of the LvNL equations for the  {\it diagonal elements}
  of the density matrix $\rho_{nn}$, when there also dissipative terms. The off-diagonal matrix
  elements $\rho_{nm} ~(n\neq m)$ will be discussed after introduction of the initial conditions.
  \beq
 \hbar\dot{\gamma}^{\alpha}_n
  = -iH_{nr}\gamma_r^{\alpha} +
       L_{{\dot n}s}\gamma_s^{\alpha}\gamma^{\alpha*}_r{L^{*}_{\dot n r}}(\gamma^{\alpha*}
       _{{\dot n}})^{-1}-
      L^{*}_{rn}L_{rs}\gamma_s^{\alpha}
 \label{gammadot3}
 \enq
 The  star denotes the complex conjugate. The summation
 convention for doubly appearing roman indices is used, but
${\dot n}$ in a subscript means no summation over $n$ (and no
summation is implied for multiple Greek indices). The inverse
$(\gamma^{\alpha*} _{{\dot n}})^{-1}$ is not an element of the
inverse matrix of $\gamma$; rather, it is the inverse of an
element of $\gamma$. For the conjugate variable one has \beq
\hbar\dot{\gamma}_n^{\alpha*}=i\gamma^{\alpha*}_r H_{rn}
     +(\gamma_{{\dot n}}^{\alpha})^{-1} {L_{\dot n r}}\gamma_r^{\alpha}\gamma_s^{\alpha*}
      L^{*}_{{\dot n}s} - \gamma_s^{\alpha*}L^{*}_{rs}L_{rn}
 \label{gammadot5}
 \enq
As discussed in previous texts  using the square-root method \cite
{Reznik,SGS}, the rate equations for $\rho_{nn}$ follow from
combination of  \er{gammadot3} and \er{gammadot5}. In fact, using
the left hand side of these expressions and carrying out the
ensemble averaging, one obtains the time-rate of a diagonal
element of the density matrix, as follows \ber \hbar \dot
\rho_{{\dot n}{\dot n}} & = &
\frac{\hbar}{A}\sum_{\alpha}[\gamma_{\dot n}^{\alpha}
{\dot\gamma}^{\alpha*}_{\dot n}+ {{\dot\gamma}}^{\alpha}_{\dot
n}\gamma^{\alpha*}_{\dot n}]
\nonumber\\
& = & \frac{1}{A}\sum_{\alpha}\{ i[\gamma_{\dot n}^{\alpha}
\gamma^{\alpha*}_r H_{r{\dot n}}-H_{{\dot n}r}\gamma_r^{\alpha}\gamma_{\dot n}^{\alpha *}]\nonumber\\
     & + & {2L_{\dot n r}}\gamma_r^{\alpha}\gamma_s^{\alpha*}
      L^{*}_{{\dot n}s} -  \gamma^{\alpha}_{{\dot n}}\gamma_s^{\alpha*}L^{*}_{rs}L_{r{\dot n}}-
      L^{*}_{r{\dot n}}L_{rs}\gamma_s^{\alpha}\gamma_{\dot n} ^{\alpha *} \}
\nonumber\\
& = & \{ i [\rho,H] +2 L \rho L^\dag-L^\dag L \rho - \rho L^\dag L \}_{{\dot n}{\dot n}}
 \label{rhodot3}
 \enr
On the right hand side all products of the $\gamma$'s an be written in terms of density matrix elements
and the expression yields precisely the $nn$- component of right hand side of \er{Lindblad}.
 Similarly
to solutions of the LvNL equations, the trace of $\rho$ is
preserved ($Tr \dot{\rho}(t)=0$ at all times) and the positivity
of any diagonal matrix elements $\rho_{nn}$ is ensured,
since $\rho_{nn}=\gamma_n\cdot\gamma^{*}_{n}\ge 0$.

On the other hand, the dissipative part of the rate expressions
for the non diagonal components of the density matrix has the form
\ber \hbar \dot \rho_{n m} & = &
\frac{\hbar}{A}\sum_{\alpha}[\gamma_{n}^{\alpha}
{\dot\gamma}^{\alpha*}_{m}+
{{\dot\gamma}}^{\alpha}_{n}\gamma^{\alpha*}_{m}]
\nonumber\\
& = & \frac{1}{A}\sum_{\alpha}\{ i[\gamma_{n}^{\alpha}
\gamma^{\alpha*}_r H_{r m}-H_{n r}\gamma_r^{\alpha}\gamma_{m}^{\alpha *}]
- \gamma^{\alpha}_{n}\gamma_s^{\alpha*}L^{*}_{rs}L_{r m}-
      L^{*}_{r n}L_{rs}\gamma_s^{\alpha}\gamma_{m} ^{\alpha *}
\nonumber\\
&+& L_{\dot n s} \gamma_{s}^{\alpha} \gamma_{r}^{\alpha *} L^{*}_{\dot n r} (\gamma_{\dot n}^{\alpha *})^{-1}
      \gamma_{m} ^{\alpha *}+ L_{\dot m r} \gamma_{r}^{\alpha} \gamma_{s}^{\alpha *} L^{*}_{\dot m s}
       (\gamma_{\dot m}^{\alpha })^{-1} \gamma_{n} ^{\alpha}
    \}
\nonumber\\
& = & \{ i [\rho,H] - L^\dag L \rho - \rho L^\dag L \}_{n m}
+ (L B^{\dot m n} L^\dag)_{\dot m \dot m} + (L B^{* \dot n m} L^\dag)_{\dot n \dot n}
\nonumber\\
&  &
 \label{rhodot4}
 \enr
in which the $B$ tensor is defined as \beq B_{rs}^{mn} =
\frac{1}{A}\sum_{\alpha} \gamma_{r}^{\alpha} \gamma_{s}^{\alpha *}
(\gamma_{m}^{\alpha})^{-1} \gamma_{n}^{\alpha} \enq This contains
terms that include the inverse quantities $
(\gamma_{n}^{\alpha})^{-1}$ and cannot be expressed in terms of
the density matrix. This was already noticed in \cite {Reznik}. In
the case that $m=n$ we obtain $B_{rs}^{\dot m \dot m} =
\rho_{rs}$. Thus, while the square root method is self consistent,
it is not fully equivalent with Lindblad
 equations. (A detailed comparison with different methods is given in section
 5.) The difference can be clearly formulated by inserting the
 following expression in the curly brackets in \er{rhodot4}
 and subtracting the same from the last two terms:\beq L^{*}_{rn}\frac{2}{A}\sum_{\alpha}
\gamma^{\alpha}_{r}\gamma_s^{\alpha*}L_{s m}=\{2L^{\dag}\rho L\}
_{nm} \label{2LrhoL}\enq Then the curly brackets becomes \beq \{i
[\rho,H] +2L^\dag \rho L - L^\dag L \rho - \rho L^\dag L \}_{n
m}\label{Lindpart}\enq which is the Lindblad expression, while the
difference can be written as
\beq\frac{1}{A}\sum_{\alpha}\gamma_s^{\alpha}(D^{mn\alpha}_{rs}+D^{nm\alpha*}_{sr})\gamma^{\alpha*}_s
\label{difference}\enq with the definition that\beq
D^{mn\alpha}_{rs}=(\gamma^{\alpha*}_n)^{-1}L_{ns}\gamma^{\alpha*}_m
L^{*}_{nr} -L^{*}_{sn}L_{rm}\label{Dfn}\enq The square-root method
leads therefore (in general) to different solutions than the
Lindblad equation. We repeat that \er{rhodot4} is not used to
obtain the off-diagonal density matrix terms; rather, these are
calculated directly from the square-root factors.
\subsubsection{Two components}
To complete this subsection we write down explicitly the equations
as they should appear for a two component system to be used in the
following sections (but suppressing the system index $\alpha$):
\ber
 \hbar\dot{\gf} & = & -i(H_{11}\gf + H_{12}\gs)
\nonumber \\
 &-& |L_{21}|^2 \gf -  L^*_{21} L_{22} \gs
 + L^*_{12} \frac{\gs^*}{\gf^*} (L_{11} \gf + L_{12} \gs)
\label {gammadotstat1}\\
 \hbar\dot{\gs} & = &
 -i(H_{21}\gf + H_{22}\gs)
\nonumber \\
 &-&  L^*_{12} L_{11} \gf- |L_{12}|^2 \gs
 + L^*_{21} \frac{\gf^*}{\gs^*} (L_{21} \gf + L_{22} \gs)
 \label {gammadotstat2}
 \enr
In the case that the Lindblad operator $L$ does not have diagonal
elements those equations can be further simplified:
\ber
 \hbar\dot{\gf} & = & -i(H_{11}\gf + H_{12}\gs)
- |L_{21}|^2 \gf + |L_{12}|^2 \frac{|\gs|^2}{\gf^*}
\label {gammadotstat1nd}\\
 \hbar\dot{\gs} & = &
 -i(H_{21}\gf + H_{22}\gs) - |L_{12}|^2 \gs + |L_{21}|^2 \frac{|\gf|^2}{\gs^*}
 \label {gammadotstat2nd}
 \enr

 \subsection {The uniqueness of the $\gb$ evolution equation}

 According to section \ref{unique} $\gb$ is not unique. $\gb$ can be replaced by
 an equally plausible representation $\gb'$ such that a unitary matrix $U$ connects
 the two:
\beq
\gb^{\alpha}_{n}(t) = \gb'^{\beta}_{n}(t) U^{\alpha}_{\beta} (t)
\label{gbp2b}
\enq
Taking the derivative of \er{gbp2b} we obtain:
\beq
\dot \gb^{\alpha}_{n}(t) = \dot \gb'^{\beta}_{n}(t) U^{\alpha}_{\beta} (t) +
 \gb'^{\beta}_{n}(t)  \dot U^{\alpha}_{\beta} (t)
\label{dgbp2b}
\enq
This can also be written as:
\beq
\dot \gb^{\alpha}_{n}(t) = U^{\alpha}_{\beta} (t)[\dot \gb'^{\beta}_{n}(t)  +
 \gb'^{\mu}_{n}(t)  \dot U^{\nu}_{\mu} (t) {(U^{-1}(t))}^{\beta}_{\nu}]
\label{dbgbp2b}
\enq
Introducing the expression from \er{gbp2b} and \er{dbgbp2b} into \er{gammadot3}
we obtain the result:
\ber
&\hbar U^{\alpha}_{\beta} (t)[\dot \gb'^{\beta}_{n}(t)  +
 \gb'^{\mu}_{n}(t)  \dot U^{\nu}_{\mu} (t) {(U^{-1}(t))}^{\beta}_{\nu}]
= U^{\dot \alpha}_{\beta} (t)[-iH_{nr}\gb'^{\beta}_{r} - L^{*}_{rn}L_{rs}\gb'^{\beta}_s&
\nonumber \\
  &+ L_{{\dot n}s}\gb'^{\beta}_s \gb'^{*\nu}_r{L^{*}_{\dot n r}} U^{*\dot \alpha}_{\nu} (t)(\gb'^{*\nu}
       _{{\dot n}}U^{*\dot \alpha}_{\nu} (t))^{-1}]&
 \label{gammadot3b}
 \enr
Multiplying by the inverse matrix of $U^{\alpha}_{\beta}$ and rearranging terms we obtain:
\ber
&\hbar \dot \gb'^{\beta}_{n}(t)= -iH_{nr}\gb'^{\beta}_{r} - L^{*}_{rn}L_{rs}\gb'^{\beta}_s
  + L_{{\dot n}s}\gb'^{\beta}_s \gb'^{*\nu}_r{L^{*}_{\dot n r}} U^{*\dot \alpha}_{\nu} (t)(\gb'^{*\nu}
       _{{\dot n}}U^{*\dot \alpha}_{\nu} (t))^{-1}&
\nonumber \\
       & - \hbar \gb'^{\mu}_{n}(t)  \dot U^{\nu}_{\mu} (t) {(U^{-1}(t))}^{\beta}_{\nu} &
 \label{gammadot4}
 \enr
We see that the third term in the right hand side of
\ern{gammadot4} (corresponding to the second term in the right
hand side of \ern{gammadot3}) is to some degree arbitrary.
Furthermore, if the matrix $U(t)$ is chosen to be time dependent a
fourth term can be added without effecting the results. How should
one choose a matrix $U(t)$? Some recommendations can be given:
\begin{enumerate}
    \item Choose a matrix $U(t)$ to simplify \ern{gammadot4}.
    \item Choose a matrix $U(t)$ to avoid singularity condition that may occur in the
    third term in the right hand side of \ern{gammadot4}. More about the problem of singular
    initial condition in the next subsection.
    \item Choose a matrix $U(t)$ in order to reduce the difference between the Lindblad formalism
    and the factorization matrix formalism. This can be done in terms of the difference tensor $D^{mn\alpha}_{rs}$
    defined in \ern{Dfn}. Inserting the expression from \er{gbp2b} into \er{Dfn} we obtain the result:
    \ber
    D^{mn\alpha}_{rs}[U^{\nu}_{\beta}]&=&
    (\gamma^{*\dot \alpha}_{\dot n})^{-1} L_{\dot n s} \gamma^{*\dot \alpha}_m L^{*}_{\dot n r} -L^{*}_{s n}L_{rm}
    \nonumber \\
    &=&
    (\gb'^{*\nu} _{{\dot n}}U^{*\dot \alpha}_{\nu} (t))^{-1} L_{\dot n s}
    (\gb'^{*\nu} _{m}U^{*\dot \alpha}_{\nu} (t)) L^{*}_{\dot n r} -L^{*}_{s n}L_{rm}
    \label{Dfn2}
    \enr
    Hence the $U(t)$ can be chosen in order to obtain smaller $D^{mn\alpha}_{rs}$ tensors.
\end{enumerate}

 For the examples which were worked out in this paper it was found that choosing the matrix $U(t)$
 as the identity matrix produced both simple equations and also good agreement with the Lindblad
 theory. However, for more involved cases a different choice of the matrix $U(t)$ may be needed.

\subsection {Initial conditions}

 The presence of the inverse in the dissipative part of the evolution equations causes
 the initial conditions (IC) to be of great importance. This is evident, because the rate
 expressions are singular whenever  a component probability is zero. We can
 simplify the treatment by expressing the density matrix at $t=0$ in a
 diagonal form. This can always be done by a suitable transformation
 of the basic states. The cases of pure and mixed systems are then easily distinguished.
 The physical IC are, quite generally, that
 \beq
 |\gamma_n^{\alpha}(0)|=\sqrt {p_n^0}~~~for~ n=1,...,M
 \label {IC1}
 \enq
 and zero for the rest of the states. Here the initial probabilities $p_n^0$
 \beq
 \sum_{n=1}^{M}{p_n^0}=1
 \label{norm1}
 \enq
 by normalization of the density matrix. In a pure state
    \beq
  M=1
  \label {pure}
  \enq
  and in a mixed state
  \beq
  M~>~1
  \label{mixed}
  \enq

We have already noted that a zero value of a $\gamma$-variable at $t=0$ (or
at any later time), causes a singularity on the right hand side of
\er{gammadot3} and  \er{gammadot5}. This is overcome by starting
the integration at a time arbitrarily close to and slightly above
$t= 0$, subject to the  $IC$'s given by
\beq
\lim _{t \to 0+}\gamma_{n}^{\alpha}(t)=
     e^{i\phi_{\alpha,n}}
     [\sqrt{p_{n}^0}+\sqrt{2{L_{{\dot n}s}\sqrt{p_{s}^0 p_{r}^{0}}~e^{i(\phi_{\alpha,s}
     -\phi_{\alpha,r})} L^{*}_{{\dot n}r}t}}~~]
  \label{IC2}
  \enq
  with the phases $\phi_{\alpha, n}$ taken to be real, but with no other restriction on them.
  When $p_{n}^0\neq 0$, the second term in the above equation can be ignored. But if the
  first term is zero, the following term is essential. The  correction  to
   this term can be shown to be, for small
  $t$, of the order  $|L|^2 t$. It can be checked that the above choice of $IC$ ensures that
   the singularity on
  the right hand side of the evolution equation for any component $\gamma$ is exactly cancelled by the singular time
  derivative on the left hand side. (For a historical remark, singularities of the solutions in the
  density matrix at $t=0$ were noted early on, in section 4 of \cite{vanHove57}. The fast
   "slippage" or initially irregular behavior of solutions was recognized in \cite{SuarezSO}.)

We now assume that all solutions of the rate equations for the
$\gamma$'s correspond to a physical system in the ensemble. The
ensemble averaging has therefore to be carried out for all
possible choices of the initial phases.
\subsubsection {Short time behavior}
Let us now calculate the time development of the density matrix at
short times, for an initially {\it pure} state, in which \beq
\rho_{11}(t=0)= 1,~ \rho_{1n}(0)=0, ~\rho_{nm}(0)=0~,~for ~n\neq 1
\neq m \label{pure2} \enq In this case \beq \sqrt{p_1^0} = 1 \enq
and \beq \sqrt{p_n^0} = 0 \qquad n \neq 1 \enq For the case $n
\neq 1$ \er{IC2} takes the form \beq \lim _{t \to
0+}\gamma_{n}^{\alpha}(t)=e^{i\phi_{\alpha,n}} \sqrt{2{L_{{\dot
n}1} L^{*}_{{\dot n}1}t}} =e^{i\phi_{\alpha,n}} |L_ n 1|\sqrt{2t}
\label{IC2b} \enq The density matrix has the following short time
behavior: \beq \lim_{t\to 0} \rho_{nm}= \lim_{t\to 0} \frac{1}{A}
\sum_{\alpha}\gamma_{n}^{\alpha}(t)\gamma_{m}^{\alpha *}(t)
\approx  g_{nm}(t) \frac{1}{A}
\sum_{\alpha}e^{i(\phi_{\alpha,n}-\phi_{\alpha,m})}~~
\label{offdiag2} \enq where \ber g_{nm}(t)& = & |L_ {m
1}|\sqrt{2t} ~ for~ n=1, m>1
\nonumber\\
& =  & 2|L_{n 1}| |L_{m 1}| t ~~for ~n\neq 1 \neq m
 \label{gnm}
 \enr
The last case includes diagonal matrix elements for the initially
absent states, $n=m>1$, but here the phase factor averaging over
$\alpha$ (the system labels) in \er{offdiag2} gives unity. These
matrix elements  are thus seen to give rise
 to a non-zero value within a time of the order of $t\approx L^{-2}$. Due to the
  dissipative mechanism (represented by $L$), the system will
 therefore become mixed beyond this time, so that  ${\bf Tr}~\rho^2<1$.
 This pure-to-mixed transition can also be obtained from calculation of the time
 derivatives of the square root factors. (With a purely
 Hamiltonian interaction ($H$), an initially pure state for this Hamiltonian will  persist
 to be a pure state at later times.)

\section {Decoherence in Molecular Systems}

\subsection {Driven Level Crossing}

 The changes in the density matrix involving  two states during fast level crossing by a molecular
 systems were considered in \cite{Kayanuma} including  also dissipative forces. It was later remarked
   in \cite{RauW} that the "square root operator
  method of \cite {Reznik}", represents an alternative way of showing how a dissipative term in the
  Hamiltonian can cause decoherence. We now
   formulate the evolution equations in this square-root scheme and  solve
   the resulting equations. The solutions for the diagonal term in the density matrix,
    shown graphically below in Figure 1 and similar
   to those shown in \cite{EnglmanY2004}, have all the
   features appearing in the earlier treatments \cite {Kayanuma,RauW} based on entirely different
    schemes of solution.

The level crossing ($LC$) system is characterized by a Hamiltonian ($H_{LC}$) and a
non - Hermitian dissipative ($L_{LC}$) part
written for the two states in the matrix forms, as follows:
 \beq
  H_{LC}=
 \left( \begin{array}{cc}
  \frac{1}{2}G \cos (\omega t) & J\\
  J & -\frac{1}{2}G \cos (\omega t)
  \end{array} \right)\label{HLC}\enq
  \beq
  L_{LC}=
 \sqrt{\Gamma}\left( \begin{array}{cc}
  0 & 1\\
  1 & 0
  \end{array} \right)
  \label{LLC}
  \enq
$G$ denotes the strength of the periodic driving
field; the coefficients $J$ and $\Gamma$ of the tunnelling and of the dissipative mechanisms
are denoted by same symbols as in \cite{Kayanuma} and \cite {RauW}.

For a Markovian process the rate equations are now written for the two
 elements $(\gamma_1^{\alpha},\gamma_2^{\alpha})$ in the density matrix as
\ber i\hbar\gd_1^{\alpha} & = &\frac{1}{2}G \cos(\omega t) \gamma_1^{\alpha} +
J\gamma_2^{\alpha} -i\Gamma[\gamma_1^{\alpha} -|\gamma_2^{\alpha}|^2 /\gamma_1^{\alpha*}]
\nonumber\\
i\hbar\gd_2^{\alpha} & = & -\frac{1}{2}G \cos(\omega t) \gamma_2^{\alpha} +
J\gamma_1^{\alpha} -i\Gamma[\gamma_2^{\alpha}  - |\gamma_1^{\alpha}|^2/\gamma_2^{\alpha*}]
\label{dissipa} \enr We have already called attention to the
divisors $\gamma^{*}$ on the extreme right, characteristic of the
factorization formalism for dissipative processes, \er
{gammadot5}. The trace of the density $\rho (=\gamma\cdot\gcr)$
stays constant during the motion, by construction.

We next solve two equations for the $\gamma's$, subject to the
pure state initial conditions $|\gamma_1^{\alpha}|^2=1$, $\gamma_2^{\alpha}\approx 0$ at
$t=0$. Then from the solutions we form the diagonal matrix elements
of the density matrix and finally show the results in figure 1. For a beginning,
the three lower drawings in the figure arose from calculations that were carried out for
  a density matrix referring
  to  an "ensemble" consisting of one system. This means that  one has
  $A=1$ and  the summation over $\alpha$ in \er {rhomatrix1} is trivial.
 The quantity
changed between the upper three drawings is the strength $\Gamma$ of the dissipative term.
 As this
increases, a transition takes place from the slow to the fast decoherence regime. We have
already noted the remarkable
   similarity of the results obtained here by the factorization
method to those in Figure 1 in  \cite{Kayanuma} and in \cite {RauW}, except that for strong
dissipation their drawings show only tiny oscillations, unlike our third drawing from below.
 In this, drawn for $\frac{\Gamma}{\omega}=20$,
 after a very steep initial downward slope (not visible in the figure), both diagonal density matrix
  elements oscillate about the asymptotic value of $\frac{1}{2}$.

 In this trivial "ensemble" of $A=1$ the off-diagonal matrix element is simply given by
  $|\rho_{12}|= |\gamma_1 \gamma_2^*|=|\gamma_1||\gamma_2|=\sqrt{\rho_{11}(1-\rho_{11})}$\footnote{
  $\rho_{11}+\rho_{22} = 1 \Rightarrow |\gamma_2| = \sqrt{\rho_{22}} = \sqrt{1-\rho_{11}}$ }.
  This differs from the same quantity calculated and shown
  in Fig. 2 of \cite {RauW}. Their curve initially oscillates around $0$, and at longer
  times settles  down to this value, whereas our result tends asymptotically to $\frac{1}{2}$.
  We do not regard this as a serious discrepancy, since in a realistic model, in which  $A$
  is large, the phase decoherence will make $\rho_{12}$ vanish, as argued in section 2.3.

   We have also calculated the density
  matrices when there is a non-trivial summation, namely when initial conditions
   are $\gamma_1^{\alpha}
(0)=e^\frac{i\pi}{2}, e^\frac{2i\pi}{2}, e^\frac{3i\pi}{2} ,
  e^\frac{4i\pi}{2}$ for $\alpha =1,2,3,4$ respectively and  $\gamma_2(0)\approx 0$,
  (so that $A=4$), instead of having only
  $\gamma_1^{\alpha}(0)=1$ (and $A=1$), as before. Physically, this represents a certain type of averaging
  over the environment. (In more
   complex systems, one would require an averaging in the density matrix over a much larger
    number of states,  such that $A \to \infty$.) The resultant density tends now to an
    almost perfectly straight line. This is similar
   to the graphs shown in both \cite{Kayanuma} and in \cite {RauW} for strong dissipation and
   elucidates the practical consequence of system-averaging in the density matrix.  We have
   also worked out the $A=4$ case for
    the lower two graphs in Figure 1. For these graphs, there was hardly any perceptible
    change from  those shown.

\begin{figure}
 \vspace{10cm}
 \includegraphics{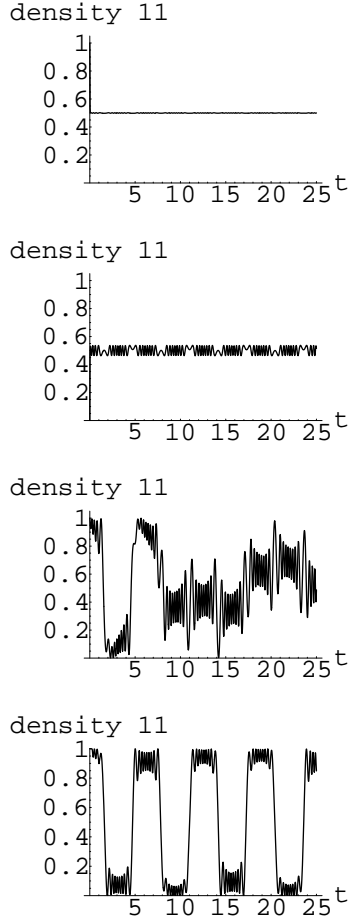}
 \caption {Evolution of a diagonal element of the density matrix $\rho_{11}$. The strength of
 dissipation increases upwards in the drawings. For all drawings the parameters in \er{dissipa} are chosen
 as $G=25$, ~$J=3$,~$\omega=1$. Then, in the bottom drawing $\Gamma$ (the dissipative
  parameter)$=0$, $A$ (the ensemble size) =$1$;
 ~in the second drawing (from below): $\Gamma=0.05$, $A=1$; in the third drawing: $\Gamma=20$, $A=1$;
  in top drawing  $\Gamma=20$ (as in previous, {\it but})~ $A=4$. The initial downward slope in
  the top two drawings is too
   steep to be visible and so are tiny fluctuations in the horizontal part of the top drawing.
   Note the smoothing effect of the ensemble averaging, evident from a comparison between the
   top two drawings.}
\end{figure}
For the physical meaning of these results in the context of molecular level crossing, we point to
 \cite{Kayanuma} and references therein, while for a more general application to decoherence
 we refer to \cite{RauW}.
\subsection{Descent Across a Conical Intersection}

Whereas in the previous case the decoherence mechanism was activated by a non-Hermitean dissipative
force, whose strength was designated by $\Gamma$, there are situations where dissipation is
intrinsic in the dynamics. A situation of this type takes place in the excited state dynamics of a
molecular system moving on a potential energy surface as it approaches a conical intersection
({\it ci}).
(This is a frequently encountered natural process, and it has been claimed that it
is basic to many naturally occurring life processes in which an electronic state is
 changed, e.g. photosynthesis \cite{Yarkony}. {\it ci} have  been previously studied by
  numerous researchers,  \cite{EnglmanYACP} and other papers in that volume \cite {BaerB}.)
 We now briefly give the underlying formal background, with a view of applying to it
 our formalism. A schematic illustration of a {\it ci} is shown in Figure 2.

 \begin{figure}
\vspace{12cm}
 \includegraphics{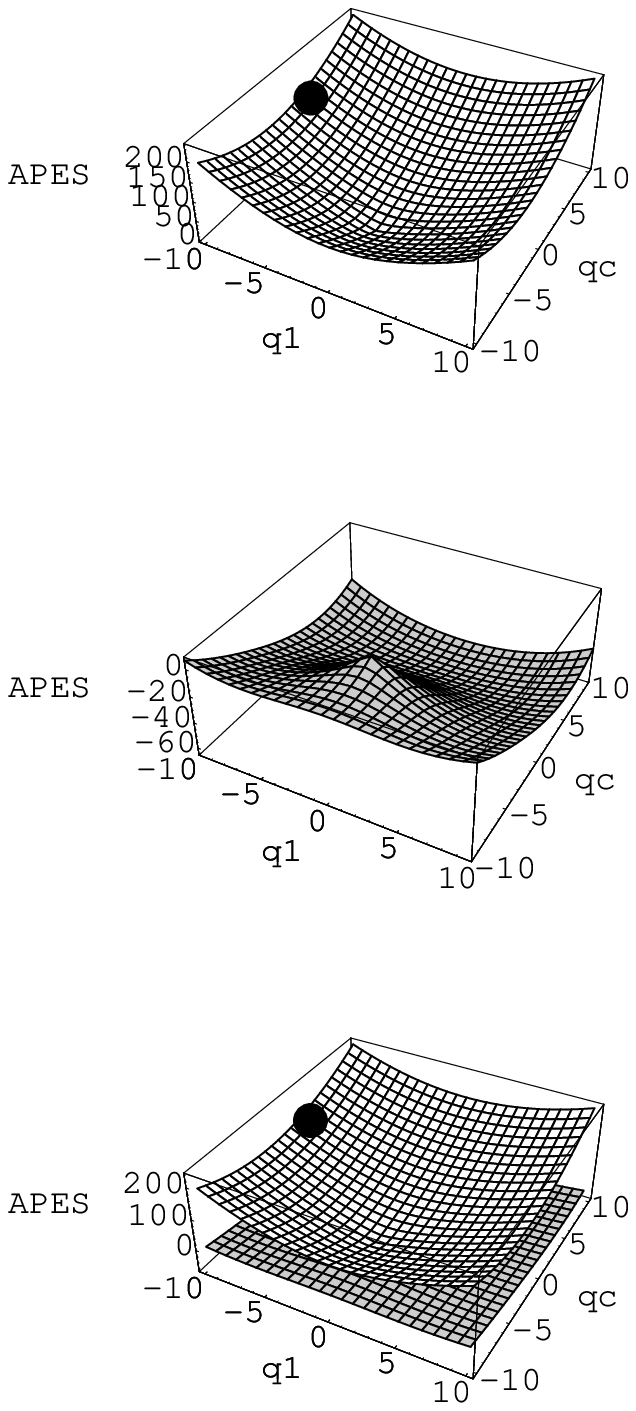}
 \caption {Schematic drawings of adiabatic potential surfaces
exhibiting conical intersection ({\it ci})
    The upper and middle drawings show the upper and lower sheets as function of
  the two displacement  coordinates $Q_1$ and $Q_c$  introduced in the
  text. The lower drawing superimposes
        these two and shows the locally conical nature of the intersection at
     $Q_1=-4$ (expressed in arbitrary units, typically about ~-0 .1 nm)  and
       $Q_c=0$. The system starts  its
        trajectory on the upper surface as a wave packet centered  at
   (say)  ($ Q_1=-10,Q_c=2$) (shown by a blob,
    where it belongs almost entirely to the upper electronic state)
  and descends towards and beyond the {\it ci},
oscillating to and forth and (partly) losing  its upper electronic state
  character.}
\end{figure}
We first recall the basic result of von Neumann and Wigner \cite {NeumannW} that the crossings
(points of degeneracy) of potential energy surfaces (=electronic energies as functions of the
nuclear coordinates) for a polyatomic molecule can be generally described in terms of just two
nuclear displacement parameters. (Following the notation of \cite {MantheK1}, we shall here denote
these  by $Q_1$ and $Q_c$, the former being a "tuning" and the latter the "coupling" mode
coordinate. They are
actually two linearly independent combinations of the nuclear coordinates.) The two surfaces
(belonging to the two locally adjacent "adiabatic" electronic states) separate from each other
 near the intersection point in a manner that is
 linear in both coordinates: hence the name "conical intersection" ({\it ci}). They differ from  energy-surface
 intersections that happen during
  molecular or atomic collisions, in that these may (approximately) be treated in a single-dimensional
  coordinate space, usually the separation between the colliding-reacting species. The
  probability of change of the electronic state in such collisions is given by the Landau-Zener
  formula \cite{Landau},\cite{Zener}, which is at the basis of the subject in the previous section.
   An analogous semi-classical formula for the passage across a {\it ci} was obtained by Nikitin \cite {Nikitinbook}.
  Later developments were summarized in \cite{DesouterDLPLL}. In a simplified form the
   expression of \cite {Nikitinbook}
 for the asymptotic probability of transfer between the two diabatic states in a $2$
 -dimensional space ($x,y$)  can be written as  \beq P^{diab}_{2\to 1}
 = \exp[-\frac{\pi}{2\hbar} \frac{K(v_x^2+v^2_y)^{-\frac{1}{2}}}{l^2}]\label{P21}\enq In
 this formula $v_x$ and $v_y$ are components of the starting velocities on the upper ("$2$")
  adiabatic surface, $l$ is the closest distance in the passage to the intersection point
  and $K$ is the strength of the  electron-nucleus dynamic coupling.

The time dependent \SE  ~for the dynamics of a {\it ci} was
subsequently solved numerically in several papers, especially in
\cite{KoppelCD} - \cite{MantheK}, which contain references to
earlier works. These show that the initial wave packet (which is
excited in an upper electronic state) has a definite, non-zero
asymptotic probability to end up in the other electronic state as
the wave packet traverses the {\it ci}. The dynamics bears
therefore the {\bf irreversibility} hallmark of a dissipative
mechanism, though such mechanism was nowhere introduced in this
model (unlike that in the previous section). The irreversibility
was explained in \cite {MantheK} as due to the essential
anharmonicity of the dynamics in the $Q_1,Q_c$-coordinates. The
anharmonicity shows up in the cusps on the energy
 surfaces at the degeneracy point.

\subsubsection {A simplified formalism for {\it ci}}

It is customary to represent the two bare or "diabatic" electronic states (those that are independent
of the nuclear coordinates) by the column vectors \beq ( "1",  "2")~=~(\left( \begin{array}{cc}
  1 \\
  0
  \end{array} \right),\left( \begin{array}{cc}
  0\\
  1
  \end{array} \right))\label {zeta}\enq
(\cite {MantheK1}- \cite {Englman}). Temporarily simplifying the situation of \cite{MantheK1}
 to  the case where there is only one tuning mode $Q_1$, as well as the coupling mode $Q_c$,
  we have for the two  coordinates the following harmonic oscillator Hamiltonian:
  \beq H_{nuc}= \frac{\hbar\omega_1}{2}(-\frac{\partial^2}{\partial Q^2_1} + Q^2_1)+
  \frac{\hbar\omega_c}{2}(-\frac{\partial^2}{\partial Q^2_c} + Q^2_c)
  \label{HO}\enq
  where $\hbar\omega_1$ and $\hbar\omega_c$ are the quanta of vibrational energies.
  In the doublet representation
   the nuclear Hamiltonian is written as a scalar or, equivalently,
  as $H_{nuc}$ times the $2$x$2$ unit matrix $I$.

   The nuclear-electronic interaction part can then be written without loss of generality as
   \beq
  H_{el-nuc}=
 \left( \begin{array}{cc}
  \delta E+K_1Q_1 & K_cQ_c\\
  K_cQ_c & - \delta E -K_1Q_1
  \end{array} \right)\label{Helnuc}\enq where $2\delta E$ is the vertical energy offset between
  the two electronic states and $K_1, K_c$ are the coupling strengths for the two coordinates.

 The \SE ~to be solved for the time dependent vector $\Psi(Q_1,Q_c,t)$ is \beq i\hbar
\frac{\partial\Psi(Q_1,Q_c,t)}{\partial t} = (H_{nucl}+H_{el-nuc})\Psi(Q_1,Q_c,t)\label{SE}
\enq subject to the initial condition at $t=0$, that $\Psi(Q_1,Q_c,0)$ is a wave packet (say, of
gaussian forms in the two coordinates $Q_1,Q_c$) on the upper potential surface, centered at some
position well above the intersection point ($Q^{{\it ci}}_1=-\frac{\delta E}{K_1},~Q_c=0$). This type of
wave packet can be conveniently formed by short duration (compared to the inverse frequencies
$\omega_1^{-1}$ and $\omega_c^{-1}$) optical excitation from some lower lying electronic state.
The wave packet will then descend towards the {\it ci} point and beyond. Throughout its
 passage the initially excited (diabatic) electronic state (say, $"1"$) will lose amplitude
 in favor of the other state "2". This loss will be especially intense near the degeneracy point.
 Ultimately, for "long" times of the order of picoseconds, the probability will tend to some
  asymptotic limit (say, $P_{1,\infty}$) between $0$ and $1$. (Figures 1-4, in \cite {MantheK1}. Actually, in any
   individual run it will slightly oscillate about the asymptotic probability.) This value
    depends on the parameters of the system and also, presumably to a lesser extent, on the initial
    conditions.

\subsubsection {A Lindblad-type formalism for {\it ci}}

In this section we present a reformulation of the previous two
state-two mode system, such that the irreversible nature of the
dynamics is built into the model (rather than emerges from the
solution). We do this by constructing a phenomenological
dissipative term in the square-root formalism, based on the
Hermitean formalism of the previous subsection. This procedure
will then lead to Lindblad-type terms in the master equations.
(Let it be emphasized that we are {\it not deriving} the Lindblad
term from a microscopic process, but are {\it formulating} the
microscopic equations in an irreversible setting for the {\it
partial}, electronic degrees of freedom, having eliminated the
nuclear coordinates.)

In the motion of the wave packet on an adiabatic surface the
electronic state amplitudes will depend on the position of the nuclear
coordinates. In the square root formalism, and relying on Appendix $A.1$,
 we write this as
 \beq
 \gamma_i^{\alpha}=\gamma_i^{\alpha}(Q_1,Q_c)~~~~~~(i=1,2)
 \label{gammai}
 \enq
  We
invert these relations to make the coordinates some functions of
the $\gamma$'s and then linearize the functional relation to be
composed of a classical (non-dissipative, real) part and a
"dissipative" part which will give rise to a Lindblad type term.
Specifically,
\beq
Q_r(t)=Q^{class}_r(t) +Q^L_{r}(t)~~~(r=1,c)
\label{qsplit}
\enq
for either coordinate and then
\ber
Q^{class}_1(t) & = & Q^0_1cos(\omega_1 t)+\frac{v_1}{\omega_1}\sin (\omega_1 t)\\
Q^{class}_c (t)& = & \frac{v_c}{\omega_c}\sin (\omega_c t)
\label{qclass}
\enr
These relations are appropriate for a classical
motion (for which the initial values of the coordinates are
$(Q^0_1,0)$ and of the velocities are $(v_1,v_c)$ ); they should
provide a fair enough description of the center of a gaussian wave
packet. Other choices for the initial values lead to similar
results for the asymptotic behavior. (But take note that the level
crossing probability is zero when the initial velocities are zero
or when the classical pathway goes across the {\it ci}. This is
evident from  Nikitin's expression for a {\it ci} level crossing
probability \cite {Nikitinbook, DesouterDLPLL}, shown above in
\er{P21}.)

The expressions for the Lindblad component of the coordinates are
more complicated and their rationale will be apparent only later.
They are \ber Q^L_{1}(t) & = & -\frac{i\Gamma}{K_1}\frac
{[(1-P_{1,\infty})|\gamma_1^{\alpha}|^2-P_{1,\infty}|\gamma_2^{\alpha}|^2](\gamma_2^{\alpha}
\gamma_1^{\alpha *}
+ \gamma_1^{\alpha} \gamma_2^{\alpha*})} {|\gamma_1^{\alpha}|^2\gamma_1^{\alpha}\gamma_2^{\alpha*}
 + |\gamma_2^{\alpha}|^2\gamma_2^{\alpha}\gamma_1^{\alpha*}}
\label{QL1}\\
Q^L_{c}(t) & = &
\frac{i\Gamma}{K_c}\frac
{[(1-P_{1,\infty})|\gamma_1^{\alpha}|^2-P_{1,\infty}|\gamma_2^{\alpha}|^2](\gamma_1^{\alpha}
\gamma_1^{\alpha*} -\gamma_2 ^{\alpha} \gamma_2^{\alpha*})}
{|\gamma_1^{\alpha}|^2\gamma_1^{\alpha}\gamma_2^{\alpha*} + |\gamma_2^{\alpha}|^2\gamma_2^{\alpha}
\gamma_1^{\alpha*}}\label{QLc}\enr Here
$P_{1,\infty}$ and $P_{2,\infty}$ are asymptotic weights or probabilities
 for the two (adiabatic) components. The time dependence
of the $\gamma$-factors is suppressed in these formula, for
brevity. Because of the irreversibility of the process, one can
assume the functions $Q_1,Q_c$ to be complex.

Substitution of the  nuclear coordinates, as given in
\er{qsplit}-\er{QLc}, into \er{Helnuc} and the use of this
$\gamma$-dependent Hamiltonian  in the first ("Hamiltonian") term
in \er{gammadot3}-\er{gammadot5}, yield after some simplification
the following Lindblad form of the rate of equations for the
$\gamma$'s \ber i\hbar\gd_1^{\alpha} & =& [\delta E+
K_1(Q^0_1\cos(\omega_1 t)+\frac{v_1}{\omega_1}\sin (\omega_1
t))]\gamma_1^{\alpha} +
 K_c\frac{v_c}{\omega_c}\sin (\omega_c t)\gamma_2^{\alpha}
 \nonumber\\
 & - & i\Gamma\frac
{((1-P_{1,\infty})|\gamma_1^{\alpha}|^2-P_{1,\infty}|\gamma_2^{\alpha}|^2)}{\gamma_1^{\alpha*}}
\label{gamci1}\\
i\hbar\gd_2^{\alpha} & = &  [-\delta E-K_1(Q^0_1 \cos(\omega_1
t)+\frac{v_1}{\omega_1}\sin (\omega_1 t))]\gamma_2^{\alpha} +
 K_c\frac{v_c}{\omega_c}\sin (\omega_c t)\gamma_1^{\alpha}
 \nonumber\\
 & + & i\Gamma\frac
{((1-P_{1,\infty})|\gamma_1|^2-P_{1,\infty}|\gamma_2|^2)}{\gamma_2^{*}}\label{gamci2}\enr
The Lindblad operator for this set of equations is:
 \beq
  L_{CI}=
 \sqrt{\Gamma}\left( \begin{array}{cc}
  0 & \sqrt{P_{1,\infty}}\\
  \sqrt{1-P_{1,\infty}} & 0
  \end{array} \right)
  \label{LCI}
  \enq
 When at $t=0$ the $"1"$-state is excited optically, the initial conditions are
 $\gamma_1(0)^{\alpha}=1, \gamma_2^{\alpha} (0)=0$ for all $\alpha$, one obtains asymptotically,
  after some short-time oscillations and rise of the occupation probability of
the non-excited "$2$"-state, \beq
\rho_{11}(\infty)=|\gamma_1(\infty)|^2=P_{1,\infty},
~~~~\rho_{22}(\infty)=|\gamma_2(\infty)|^2=P_{2,\infty}=1-
P_{1,\infty} \label{asymp} \enq The behavior of $\rho_{11}$ as
function
 of time is shown in figure \ref{fig3}. The parameters $K_1,K_c,\Gamma,\omega_1, \omega_c$ are all functions
 of the parameters in the molecular Hamiltonian (\er{HO},~\er{Helnuc}). $Q^0_1, v_1, v_c$ are
  defined by the mode of excitation, $P_{1,\infty}$ depends on the molecular Hamiltonian and,
  as shown in section VI of \cite {MantheK}, approximately  by the potential surfaces.

\begin{figure}
\vspace{5cm}
\includegraphics{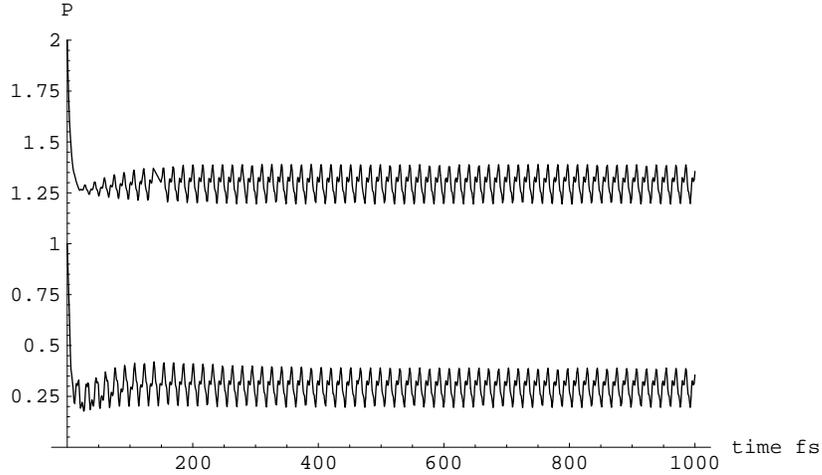}
 \caption{
Time development of the weight of the diabatic-state,
$|\gamma_1^{\alpha}(t)|^2$, following excitation into state $"1"$ and
during a passage across a conical intersection with a lower lying
state ($"2"$). Lower curve: a single run. Upper curve (displaced
upward by 1): ensemble averaging over four runs ($A=4$, explained in
text). The parameters in \er{gamci1} are ( with frequencies  and
energies $/\hbar$  in inverse femtosecond units, lengths in units
of zero point motion amplitudes, typically .05 nanometers ):
Energy offset, $\delta E =0.2$,~Coupling constants:~$  K_1
=0.06,~K_c=0.05$. Mode
frequencies:$~\omega_1=\omega_c=0.4$.~Initial velocities:
$v_1=0.24, v_c=0.32$. Dissipative strength, $\Gamma=-0.1$;
Asymptotic weight: $P_{1,\infty}=0.25$}
\label{fig3}
\end{figure}

It is proper to characterize the present computed case as
belonging to the {\it moderately strong} dissipative case, since
the oscillations about the asymptotic value begin after a time of
the order of a vibrational period . A closer look at the curves
    shows that the computed  mean asymptotic line lies above the  "input"
    asymptotic weight  $P_{1,\infty}=.25$ by about 0.03. This is an interesting effect,
    likely to be due to oscillations  in  the asymptotic  probability which arise from the
    Hamiltonian part of the rate \er{gamci1}. (We call attention to the  discussion
     in section  VI  of \cite{MantheK1}, concerning the discrepancy of about the same magnitude
      and sense  between their classical and  computed asymptotic  probabilities.) Trivially,
      the other diagonal matrix ($\rho_{22}$) tends to a value that is {\it lower} than its
      asymptotic value, since the two diagonal matrix elements add up to unity.
        The fluctuations have a roughly uniform period of $2\pi/{\omega_1}$.

        The lower curve is for a single run ($A=1$), the upper curve (vertically
        displaced for
        clarity by unity) is after "ensemble averaging" over four runs ($A=4$), which
         differ from each other by having different  phases
         ($\phi_1^{\alpha} =0$, $\pi/2$, $\pi$, $3\pi/2$, for $\alpha=1,..,4$) in
          the initial $\gamma_1^{\alpha}(0)$.
          As seen in Figure 3, no qualitative difference is observed between the curves for a
           single system and those for the ensemble averaged density matrix,
           showing that the single  system fluctuations due to the periodic
            Hamiltonian term are {\it not} averaged  out by the dissipative term.

 This last finding changes radically when we apply our formalism to a more complex
  case, that was proposed in  \cite {MantheK1} as representative of a $C_2H_4^+$ molecular
  system. There are now two tuning modes (designated by the subscripts $1$ and $2$), as well
   as different frequencies for the three modes. The results for the diagonal density matrix
   are  shown in Figure 4.
\begin{figure}
 \vspace{5cm}
\includegraphics{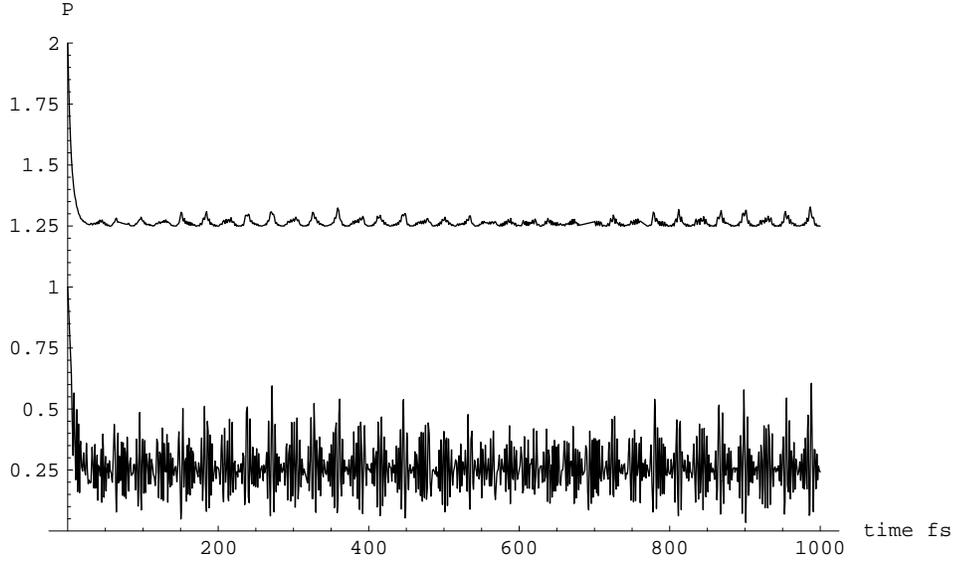}
 \caption {
 Computations in a model for $C_2H_4^{+}$
 (for a single run, the lower curve, and  with an "$A=4$ ensemble averaging" as in
  the previous figure,
 the upper curve displaced by unity for clarity).
The parameters of Table I (A) and Figure 1  (a) of \cite  {MantheK1} are
  used with  added values for initial velocities and asymptotic probabilities
    (chosen somewhat arbitrarily),  as follows:
$\delta E =0.95$,~Coupling constants:$  K_1 =0.19, K_2=-0.27
,~K_c=0.5$, mode frequencies:$~\omega_1=0.36,\omega_2=0.21,
\omega_c=0.11$,~velocities: $v_1=.6,v_2=0, v_c=.8$,
 $~ \Gamma=-.1$ (in units as in the previous figure);
      asymptotic weight: $P_{1,\infty}=0.25$}
 \end{figure}

   The fluctuations are here considerably more congested  than in the preceding
 case, where there was only a single mode-frequency.  One sees in Figure 3 that the
 dissipative mechanism does not completely eliminate the fluctuations. However, additional
 computations (not shown here) indicate that increasing the dissipation strength $\Gamma$
 does further slightly smoothen the intrinsic fluctuations. On the other hand, in the ensemble
  averaged probabilities (the upper curve in Figure 4)
  the fluctuations due to the quasi-periodic Hamiltonian are almost completely washed out.
 We finally note that the  computed  mean asymptotic  value still lies about 0.03
  above the nominal asymptotic weight.

\section {Some Models of Dissipative Processes}
\subsection{Non-Markovian processes with memory}

Following a work by Diosi et al. \cite{DiosiGS}, we consider the
situation in which the Lindblad operator $L$ is dependent on time
and hence has the faculty of memory. We consider the simple case
of a two-level system, in which: \ber H &=& \frac{1}{2} w \sigma_z
\nonumber \\
L &=& \sqrt{ f(t)} \sigma_{-}
\enr
The memory function is given by:
\beq
f(t) = \frac{G_1}{2} - \frac{\sqrt{G_1^2 - 2 G_1 l^2}}{2}
\tanh[\frac{1}{2}t\sqrt{G_1^2 - 2 G_1 l^2} +
          {\rm arctanh}[\frac{G_1}{\sqrt{G_1^2 - 2 G_1 l^2}}]]\label{memory}
\enq
The above Hamiltonian and Lindblad operator lead to the equations:
\ber
{\dot \gamma_1^{\alpha}}  &=& -\frac{1}{2}i w \gamma_1^{\alpha} - f(t) \gamma_1^{\alpha}
\nonumber \\
{\dot \gamma_2^{\alpha}}  &=& \frac{1}{2}i w \gamma_2^{\alpha} + f(t)
\frac{|\gamma_1^{\alpha}|^2}{\gamma^{\alpha*}_2} \enr The time evolution of the
expectation value of the Pauli matrices is plotted in Figure
\ref{memoryfig} in the case that the above differential equations
are solved for the pure state initial conditions $\gamma_1^{\alpha}
(0)=\frac{3}{\sqrt{10}},\gamma_2^{\alpha} (0)=\frac{1}{\sqrt{10}}$ independent of $\alpha$. The
obtained results are very similar to those of \cite{DiosiGS}.
\begin{figure}
\vspace{5cm}
\includegraphics{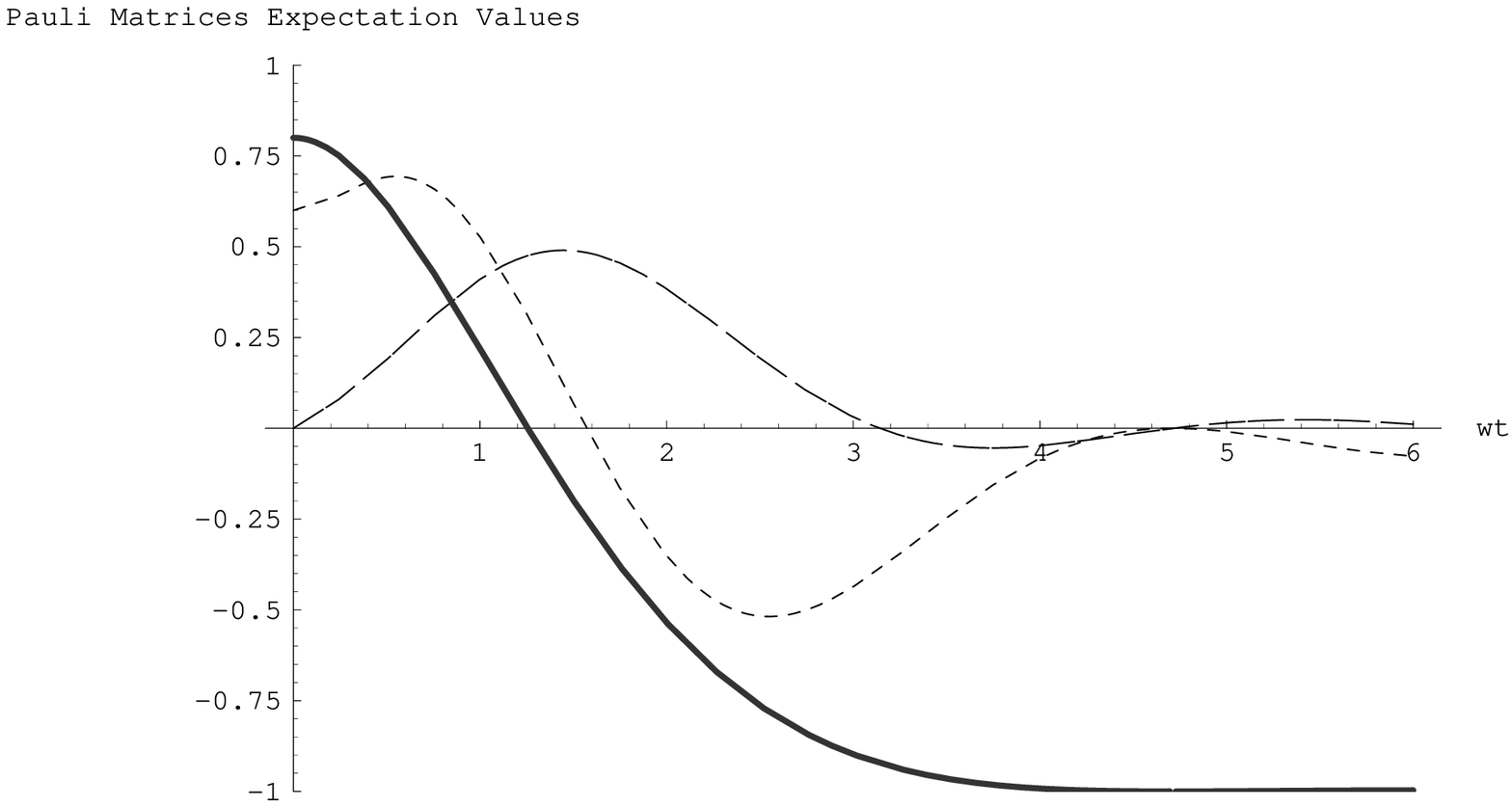}
 \caption {Expectation values of Pauli matrices for a two - level system undergoing  non -Markovian
 quantum diffusion (NMQD) for parameters as in Fig. 2(b) of \cite{DiosiGS} (w = 1; L = 1; G = 1; W1 = 1; G1 = G; a = 0).
Full  line : $ = < \sigma_z > $; short, broken lines : $ = < \sigma_x > $;
long - broken lines : $ = < \sigma_y > $}
\label{memoryfig}
\end{figure}

\subsection {Probabilities for a two-state system coupled to a bosonic reservoir}
This problem was recently studied, alongside with other instances also involving
memory, in \cite {Breuer1, Breuer2}. The case that we numerically solve using the square-root
 formalism possesses an exact solution [shown in Eq. 57 of \cite {Breuer1}], with which our
 computed values agree perfectly (Figure 7).

In the model a degenerate two-state system, $|1>$ and $|2>$ (whose
Hamiltonian is taken as zero), is coupled to a reservoir of bosons
through a time-dependent Hamiltonian interaction term. This is
given by \beq H_{int}= B(t)|2><1| + B^{*}(t)|1><2|
 \label{Hint}\enq where $B^{*}(t)$ is a reservoir excitation operator [Eq.(52) in \cite {Breuer1}]. The essential quantity
 in the dynamics is the spectral density given by \beq J(\omega)= \frac{1}{2\pi}\frac
 {l G_1^2}{(\omega-\omega_0)^2+ G_1^2)}\label {J}\enq  where $l$ is the interaction
 strength, $G_1$ is the spectral width and  $\omega_0$ the characteristic frequency in
 the reservoir. The analytical solution for the  probability of the state occupation $|2>$
  that is excited at $t=0$, is \beq \rho_{22}(t)=e^{-G_1 t} [\cosh\frac{\Omega t}{2}+
  \frac{G_1}{\Omega}\sinh\frac{\Omega t}{2}]\label{prob2}\enq where $\Omega= \sqrt{G_1^2
  -2l^2G_1}$.

  In our formalism, we solve numerically the master equations for $\gamma_1^{\alpha}(t)$,
  $\gamma_2^{\alpha}(t)$ and their complex
  conjugates, employing a Lindblad operator $L= f(t)|2><1|$, with the memory function $f(t)$
   given in \er{memory}. Initial conditions are the same for all\footnote{This means that
   the ensemble contains only a single system}  $\alpha$. The result is
    shown in Figure 6, which completely overlaps the exact
    result from  \er{prob2}. The Monte Carlo method of \cite{Breuer1,Breuer2} also agrees
   with the analytic results, but uses $10^7$ runs. It is not clear to us what is  the
   minimum runs that is required to obtain agreement, but the expediency of the square root
   method is evident.

\begin{figure}
\vspace{5cm}
\includegraphics{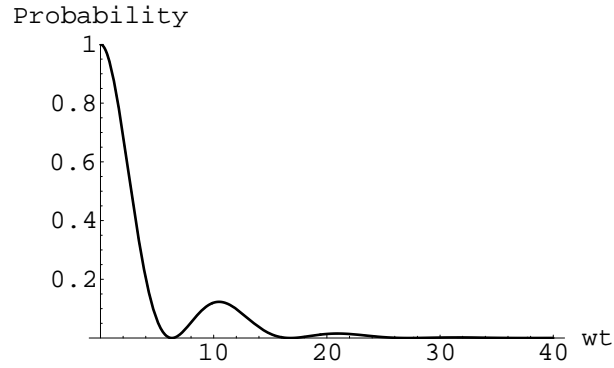}
 \caption {Probability  of excited state occupancy $\rho_{22}(t)=
  |\gamma_2^{\alpha}(t)|^2$  against normalized time ($\omega t$).
  The parameters are $l=1$, $G_1=0.2$ . Results are computed by the square root
    method. They overlap completely the analytic expression.}
 \end{figure}
In a further figure we also show the expectation value of the three angular momentum
matrices taken with respect to the state amplitudes ($\gamma_1^{\alpha}(t),\gamma_2^{\alpha}(t)$).
(Figure 7). They are similar to the drawings in Figure 6, but with more persistent
 oscillations.
 \begin{figure}
\vspace{8cm} \includegraphics{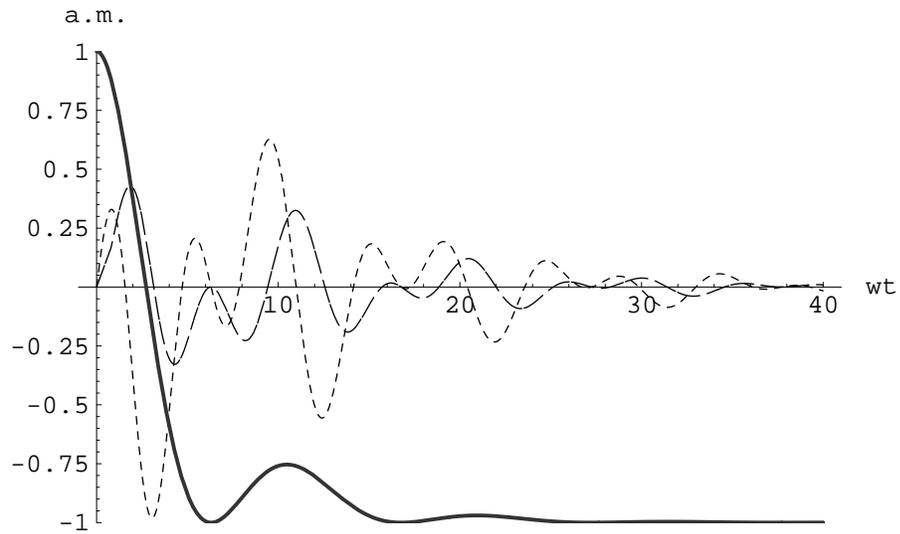}
 \caption {Averages of Pauli, angular momentum matrices for the solutions against reduced time.
$<\sigma_z>$: full lines;~$<\sigma_x>$: long broken
lines;~$<\sigma_x>$: dashed lines. Parameters  are as in Fig. 6}
 \end{figure}

\section {Comparison with Other Methods.}
In this section we briefly consider similarities and differences between the "square-root
method" and some other methods, already mentioned in the Introductory section.

(I)Number of independent rate equations.

At any instance of time the density matrix, whether by its definition in \er{rhomatrix1} or
by environment tracing , is composed
of $N$ x $A$ independent (complex) numbers ($N$ being the nominal number of states and
$A$ either the number of systems in the ensemble or the total number of states in the
environment). Thus, in principle, one needs just $N$ x $A$ rate equations to
obtain the density. The square root method postulates $N$ x $A$ equations. However,
 the roles of $N$ and $A$ are different and we find that frequently one can get
good approximations (to the important diagonal terms) by keeping $A=1$ or another small number.
(e.g., $A=4$). (Non-diagonal terms may need larger $A$-values)

Rate equation methods based on the density matrix directly work in principle with
$\frac{1}{2}N(N+1)$ equations (cf. \cite{RauW} and references), which number is for
high values of $N$ much larger than that needed in the square root method. The frequently used
Monte Carlo or "unravelling" method solves $N$ x $M$ equations, where $M$ brings in the
stochasticity of the environment
and can be very  large , e.g. 100 in \cite {GisinP}, 4096 in \cite {Goa}and $10^7$ in \cite{Breuer1}).

(II)  A New Form of the Rate Equations.

The square-root
 method is formally  unlike previously employed methods, such as those surveyed in
 \cite {KohenMT}, and it can be reasonably expected not to be reducible to them.
 To support this claim,  we call attention to the {\it inverse} "square-root factors"
  $\gamma^{-1}$ or $(\gamma^+)^{-1}$ appearing in \er{gammadot3} and other equations that
  follow on. (These factors possess an apparent similarity with the "continued fraction method" described in
  \cite {Risken} and recently applied in \cite {GarciaP}, but are essentially different.)
  The presence of these inverse factors finds expression also in a novel form of the initial
  conditions, introduced in section 2.1.

    (III) Initial time randomization

  As has been shown in section 2.1, the environment induced
  randomization affects the $\gamma$-variables only at the starting time, not during the
  time evolution. This approach was stressed in two early papers by van Hove \cite
  {vanHove55,vanHove57} in connection with the derivation of the Pauli master equation. In our
  formulation  the initially randomized $\gamma$-s lead,  in turn, to the ensemble averaging over various phases
  in a natural way. Thus the square-root procedure appears to be self  consistent.

  In the unravelling method, randomization takes place throughout the time development
  (namely, by the action of random forces on the system). It seems to us that the ensemble
   averaging
  as done in the square root method (namely, by introducing randomness only at the beginning) is
  closer to the meaning originally ascribed to the density matrix. As asserted at various
  places in \cite {vanHove55}, not only is the repeated stochasticity "superfluous", but
  "it is strictly speaking incompatible with the Schr\"odinger equation". (We understand
  that van Hove refers here to the form of the Schr\"odinger equation as a first order time-
  dependent differential equation, so that the system's evolution is fully determined by
  the initial conditions.)

(IV) Form of the Lindblad operator

As emphasized at several points in this paper, the choice of the
Lindblad operator $L$ in the square root method  is determined by
the requirement that the evolution equation for the $\gamma$'s
should lead to  rate  equations for the diagonal terms in the
density matrix, that are formally identical to those in Lindblad's
method \cite{Lindblad}. (We assume that this choice is unique.)
Thus, we do not derive the $L$-operator from an atomic model for
the system and its environment. In this sense, the square root
method is "phenomenological".

However, most density matrix methods are such, too. As an example,
we call attention to the cases studied in the classic paper of
\cite{GisinP}. In some works, such as \cite{Diosi} on quantum
diffusion, a Lindblad operator was derived from a kinematic model,
but only through making some approximations. We have already noted
some dissension, \cite {Weiss}-\cite{RomeroTH}, regarding  the
employment of the Lindblad formalism, especially for  initial
conditions that are non-separable between the system and its
environment \cite{Pechukas,RomeroTH}.

(V) Pure-to- mixed state transition

It has been shown in section 2.1 that in the presence of dissipative processes, a pure state
 becomes a mixed state in a time of
order $L^{-2}$. This is also the standard result in other formalisms, obtained theoretically,
e.g., in \cite{SuarezSO}, and by numerical calculations \cite{RauW}.

(VI) Off-diagonal terms in the density matrix.

For dissipative systems here lies probably the most significant discrepancy between the square-root
and density matrix methods. (In a Hamiltonian system, the two formalisms are equivalent, as already
noted.)  Our defining equations for off-diagonal elements differ from Lindblad's equations, and
 therefore so do (in general) the calculated values of the diagonal terms. In a particular
 case, "level crossing" treated in section 3.1, we have pointed out the discrepancy involved in the square-root
 procedure. Even in this case, the calculated behavior of the diagonal element was very similar to that
 obtained by another method in \cite {RauW}.

 It seems that the source of the discrepancy is in the different approximations made to
 arrive at a dynamical
 semi-group from a Hamiltonian system, either by truncation of the time domain or by the handling
 of phase decoherences \cite {Lindblad}. So far we have not found any result that
 would tend to invalidate the present approach, or even to lay bare any shortcomings.
 In view of the questions surrounding the positivity of linear maps with initially entangled
 states \cite {JordanSS},
  the circumstance that our rate equations for non-diagonal terms differ from Lindblad's
should not invalidate the square root method.

\section {Conclusion}

The "square-root"  method (previously used to minimize
 the action in a general time dependent process \cite{EnglmanY2004}) has now been
applied to several situations where decoherence is expressible by a Lindblad formalism.

Two irreversible molecular processes (a driven Landau-Zener process and the descent
to equilibrium across a conical intersection) were formulated by the square root
formalism.
Further illustrations of the method were quantum
diffusion and memory processes. When compared with results given in the
 literature and obtained using different procedures, the present method has led in all
 cases to good agreement and at the cost of very much less effort. The success of the method
 would seem to justify future uses of the formalism, as for Quantum Brownian Motion,
already widely studied in the literature with density matrix methods, \cite {Englman81}
-\cite{VanF} and regarding the question of thermalization \cite{AlickiL, Nest}.

At the same time, since the basic equations in the square root method and other approaches
 appear to be different in some respects, there  remains a theoretical challenge to explain
  the agreement between the results. Another task is the justification of the basic
  rate-equation expressions in the formalism by starting from microscopic models.  \\

{\LARGE \bf Appendix}

\appendix

\section {The square-root factored density matrix.}
\subsection {Matrix formulation of square root factors}
Though the density matrix in the square root factor form has
 already appeared before in \cite {Reznik} and \cite {SGS}, we introduce it
  here following textbook descriptions of von Neumann's matrix
  method (\cite {Band},\cite{Neumann}).
  Let $\Psi_{\alpha}$ be a possible wave function describing the quantum state
   of the $\alpha$'th system in the ensemble $(\alpha=1,2...A)$. It
   can be expanded in terms of an orthonormal set of eigenstates $u_n$ as
   \beq
   \Psi_{\alpha} = \sum_{n=1}^N \gamma^{\alpha}_n u_n
   \label {expansion}
   \enq
   Here the size of the orthonormal set (in principle, infinite) is taken
   as having a finite size $N$.
   From the normalization of all the wave functions $ \Psi_{\alpha}$
   we obtain:
    \ber
   1 &=& \langle\Psi_{\alpha}|\Psi_{\alpha}\rangle = \langle\sum_{n=1}^N \gamma^{\alpha}_n u_n|
   \sum_{m=1}^N \gamma^{\alpha}_m u_m\rangle
\nonumber \\
   &=&
   \sum_{n=1}^N  \sum_{m=1}^N \gamma^{\alpha*}_n \gamma^{\alpha}_m \langle u_n|u_m\rangle=
     \sum_{n=1}^N  \sum_{m=1}^N \gamma^{\alpha*}_n \gamma^{\alpha}_m \delta_{nm}=
 \sum_{n=1}^N |\gamma^{\alpha}_n|^2
   \label {expansion2}
   \enr
   As derived in \cite {Band} and other texts, the density matrix arises
   from the ensemble average over all systems in the sense that its
   $nm$ component is
   \beq
   \rho_{nm} = \frac {1}{A} \sum
   _{\alpha=1}^A \gamma^{\alpha}_n\gamma^{\alpha *}_m
   \label {rhomatrix1a}
   \enq
   The $\gamma^{\alpha}$'s are rectangular matrices of size $N$ x $A$, distinct for each
   $\alpha$ (or system in the ensemble) and the $\gamma^{\alpha *}$'s are conjugate matrices of size
   $A$ x $N$.   Calculating the trace of the matrix $\rho$ we obtain:
   \beq
   {\rm Tr \ } \rho= \sum_{n=1}^N \rho_{nn}= \sum_{n=1}^N \frac {1}{A} \sum
   _{\alpha=1}^A \gamma^{\alpha}_n\gamma^{\alpha *}_n =
    \frac {1}{A} \sum_{\alpha=1}^A 1 = 1
   \label {rhomatrix1b}
   \enq
for which we have used \erb{expansion2}.

\subsection {Properties of the density matrix}

\subsubsection {The density matrix eigenvalues}

Let us derive some of density matrix well known properties.
First $\rho$ is hermitian, since $\rho^\dag = \rho$, as can be
clearly seen from the definition of $\rho$ given in \er{rhomatrix1}.
Thus, all eigenvalues $\lambda_i$ of $\rho$ are real which is a property of all
hermitian matrices.

Second, the sum of all the eigenvalues is unity. This can easily
be shown as follows. There exists a basis of eigenvectors in which
$\rho$ is diagonal: in this basis $\rho$ will be denoted by $\rho_D$.
The diagonal elements of $\rho_D$ are just the eigenvalues $\lambda_n$.
Furthermore, given $\rho$ in an arbitrary basis
there exists a unitary matrix $U$ which transforms $\rho$ to $\rho_D$
according to the following equation:
\beq
\rho_D = U \rho U^\dag \Rightarrow \rho = U^\dag \rho_D U
\enq
By virtue of \er{rhomatrix1b} we obtain:
 \beq
   1={\rm Tr \ } \rho= {\rm Tr \ } U^\dag \rho_D U = {\rm Tr \ } \rho_D
   =\sum_{n=1}^N \lambda_n
 \label {eigsum}
 \enq

Third, we will show that all $\lambda_n$ are positive (or zero).
To prove this take an arbitrary vector $X$ one can see that: \beq
X^\dag \rho X  = \sum_{n=1}^N \sum_{m=1}^N \frac {1}{A}
\sum_{\alpha=1}^A X_n^* \gamma^{\alpha}_n \gamma^{\alpha *}_m X_m
\enq In which the definition of $\rho$ given by \er{rhomatrix1}
was utilized. Next define $R^{\alpha} = \sum_{n=1}^N X_n^*
\gamma^{\alpha}_n$, and obtain the result: \beq X^\dag \rho X =
\frac {1}{A} \sum_{\alpha=1}^A |R^{\alpha}|^2 \geq 0 \enq Further,
we write $X$ in the eigenvector basis $V_i$ as $X= \sum_i C_i V_i$
which yields: \beq X^\dag \rho X =\sum_i C^*_i V^\dag_i \rho
\sum_j C_j V_j =\sum_i C^*_i V^\dag_i \cdot \sum_j \lambda_j C_j
V_j \enq but since $V^\dag_i \cdot V_j = \delta_{ij}$ we have:
\beq \sum_i \lambda_i |C_i|^2 =  X^\dag \rho X \geq 0 \enq and,
since the $|C_i|$'s are arbitrary, we obtain $\lambda_i \geq 0$
for every $i$. From the positivity of $\lambda_i$ and \er{eigsum},
we reach the main conclusion of this section, that is: \beq 0 \leq
\lambda_i \leq 1  \qquad \forall i \label{eigbounds} \enq

\subsubsection {Pure and mixed states}

From \er{eigsum} and \er{eigbounds} one can reach the following
classification of density matrices: either there exists a special
index $s$ such as $\lambda_s = 1$ while for all $i \neq s$
$\lambda_i = 0$ or that $\lambda_i < 1$ for all indices $i$. The
first case is denoted as a "pure" state while the second is
denoted as a "mixed" state.

For the pure state we have in the diagonal basis $\rho_D^2= \rho_D \rho_D = \rho_D$
and also in an arbitrary basis
\beq
\rho^2 = \rho\rho = U^\dag \rho_D U U^\dag \rho_D U = U^\dag \rho_D \rho_D U = U^\dag \rho_D  U
=\rho
\enq
which is a necessary and sufficient condition for a density matrix to describe a pure state.
One obviously obtains also the result
\beq
{\rm Tr \ } \rho^2 = {\rm Tr \ } \rho = 1
\enq
For the mixed state we have
\beq
{\rm Tr \ } \rho^2 = {\rm Tr \ } \rho_D^2 = \sum \lambda_i^2 < \sum \lambda_i = {\rm Tr \ } \rho = 1
\enq
In summary, we conclude that
\beq
{\rm Tr \ } \rho^2 \leq 1
\enq
in which the equality sign is appropriate only in the pure case.

As an example for a pure state take an ensemble for which $A=1$
and $\rho_{nm} = \gamma_n\gamma^*_m$ in this case:
\ber
{\rm Tr \ } \rho^2 &=& \sum_n (\rho \rho)_{nn} =\sum_n \sum_m \rho_{nm}\rho_{mn} =
\sum_n \sum_m \gamma_n\gamma^*_m \gamma_m \gamma^*_n
\nonumber \\
&=& \sum_n |\gamma_n|^2 \sum_m |\gamma_m|^2 =1
\enr
Another obvious case of a pure state is an ensemble with an arbitrary number $A$ of
wave functions , but in which all the wave functions are equal. Since
$\rho_{nm} = \frac {1}{A} \sum_{\alpha=1}^A \gamma^{\alpha}_n\gamma^{\alpha *}_m
 = \gamma^{1}_n\gamma^{1 *}_m$ the same argument as above can be applied.

It remains to show that in case that not all the wave function are equal
(in a non trivial sense) we obtain ${\rm Tr \ } \rho^2 < 1$.

\subsubsection {Mixed states}

Let us calculate the trace of the square density matrix:
\beq
{\rm Tr \ } \rho^2  = \sum_n \sum_m \rho_{nm}\rho_{mn} =
  \sum_n \sum_m \frac {1}{A} \sum_{\alpha=1}^A \gamma^{\alpha}_n\gamma^{\alpha *}_m
  \frac {1}{A} \sum_{\beta=1}^A \gamma^{\beta}_m\gamma^{\beta *}_n
\enq
By defining the "state averaged density function" by:
\beq
M_{\alpha \beta} = \sum_n \gamma^{\alpha}_n \gamma^{\beta *}_n
\enq
we see that:
\beq
{\rm Tr \ } \rho^2  = \frac {1}{A^2} \sum_{\alpha=1}^A \sum_{\beta=1}^A |M_{\alpha \beta}|^2
\enq
Hence we need to show that for all $\alpha$ and $\beta$, $ M_{\alpha \beta}\leq 1$
we shall be particular interested in the case that the inequality is definite
that is $ M_{\alpha \beta} < 1$.

Let us look at two arbitrary complex vectors: $F,B$ and a complex scalar
$\alpha = |\alpha|e^{i \phi}$. Obviously
\beq
\sum_i |F_i + \alpha B_i|^2 \geq 0
\label{ineq1}
\enq
However
\beq
\sum_i |F_i + \alpha B_i|^2  = \sum_i |F_i|^2 + |\alpha|^2  \sum_i |B_i|^2+
\alpha \sum_i B_i F_i^* + \alpha^* \sum_i B_i^* F_i
\enq
Now denote
\beq
a = \sum_i |B_i|^2, c=\sum_i |F_i|^2
\enq
which are both real and positive quantities, and
\beq
2b = e^{i \phi} \sum_i B_i F_i^* + e^{-i \phi} \sum_i B_i^* F_i
\enq
which is a real quantity. And we arrive at the inequality
\beq
a |\alpha|^2 + 2b |\alpha| + c \geq 0
\enq
Since, as function of $|\alpha|$, this is an equation of a parabola  which has
all its values above the $|\alpha|$ axis (except
when the equality holds in this case the parabola touches the
axis in a single point), it follows that the
discriminant of this equation is either negative or zero, the latter in
the case that the parabola touches the axis in a single point.
Thus we obtain
\beq
b^2 \leq ac
\enq
Now assume that the length of the $F$ and the $B$ vectors is unity.
That is $a=c=1$. Further more denote $X=\sum_i B_i F_i^*=|X|e^{i\varphi}$.
Thus $b$ is equal to
\beq
b=|X| \cos(\phi+\varphi)
\enq
The following inequality is obtained
\beq
|X|^2 \leq \frac{1}{\cos^2 (\phi+\varphi)}
\enq
By choosing the arbitrary phase
\beq
\phi = -\varphi \pm n \pi
\enq
we obtain the result
\beq
|\sum_i B_i F_i^*|^2 = |X|^2 \leq 1
\label{enq3}
\enq
this being a special case of the Cauchy-Schwarz inequality. Next let us discuss
the case in which the equality sign holds
in the above equation, that is the case in which $b^2 = ac$ and the equation
$a |\alpha|^2 + 2b |\alpha| + c = 0$ is satisfied for a single value
of $|\alpha|$. This can be traced to \er{ineq1} for which we have
\beq
\sum_i |F_i + \alpha B_i|^2 = 0
\label{ineq1b}
\enq
but this is only possible for $F_i =- \alpha B_i$.
However, since the length of both vectors is $1$ we arrive
at the equation
\beq
F_i =- e^{i \phi}  B_i = e^{i (\phi+\pi)}  B_i
\enq
Thus in order for the equality to hold the vectors $F$ and $B$
must be the same up to a "global" phase. We can now show
that $ M_{\alpha \beta}\leq 1$ and in particular $ M_{\alpha \beta} < 1$,~
if $\gamma^{\alpha}$ and $\gamma^{\beta}$ are different in a "non-trivial" way
(a global phase change does not count as a difference).
This is done by identifying $F=\gamma^{\alpha}, B=\gamma^{\beta}$.
Thus, according to \er{enq3},
\beq
|M_{\alpha \beta}|^2 = |\sum_n \gamma^{\alpha}_n \gamma^{\beta *}_n|\leq 1
\enq
and, in particular,
\beq
|M_{\alpha \beta}|^2 = |\sum_n \gamma^{\alpha}_n \gamma^{\beta *}_n|< 1
\enq
unless $\gamma_n^{\alpha}$ and $\gamma_n^{\beta}$ are the same up to a global phase.
This yields, barring a unique case,
\beq
{\rm Tr \ } \rho^2  = \frac {1}{A^2} \sum_{\alpha=1}^A \sum_{\beta=1}^A |M_{\alpha \beta}|^2<1
\enq
Thus a mixed state is really mixed in the sense, that is should contain at least
two wave functions which are different in a non trivial way.

\end {document}